**Investigating the Impact of US Sovereign Credit Rating Downgrade on the US Equity Market**

**Japheth Torsar Jev, M.Sc, ACA, CIMA, CGMA**

**Lecturer in Accounting and Finance**

**School of Business and Law**

**Birmingham Newman University**

**United Kingdome**

**Email: Japheth.jev@staff.newman.ac.uk**

## 1.0 Introduction

The importance of credit ratings assigned to sovereign countries by Credit Rating Agencies (CRAs) cannot be overemphasized as they not only open access to global capital markets but also determine the cost of borrowing for countries (Jeanneau & Mukherjee, 2013; and Caballero et al., 2008) ). Following the emergence of countries from the COVID-19 pandemic, investors and creditors demand for information about countries' creditworthiness and the probability of failure to repay their debts has increased considerably. This is a result of the structural shift in the economy as well as the fiscal and monetary policy response of nations to the pandemic that has heightened their sovereign risk profiles. Thus investors need deeper insights from sovereign credit ratings to reassess countries' fiscal consolidation plans and how they plan to repay their debts during the post-Covid-19 era as those debt levels have risen to unprecedented heights (Jeanneau & Mukherjee, 2013). Credit Rating Agencies (CRAs) such as Moody's Investors Services, Standard and Poors, and Fitch Ratings adopt vigorous methodologies to collect and process both quantitative and qualitative data from sovereign countries based on which they assign credit ratings that show the probability that a country will repay both interest and principal amounts they have borrowed as and when due (Caballero et al., 2008).

Recent structural shifts in economic policies and conditions of countries caused by the COVID-19 pandemic have prompted CRAs to review the Sovereign credit ratings of most countries – a situation that has resulted in significant changes to the sovereign credit ratings of many countries (Bekaert et al., 2021). On the 1$^{st}$ of August 2023, Fitch Ratings downgraded the sovereign credit rating of the United States government from the highest possible rating of AAA to a lower possible rating of AA+. The AAA rating is the highest possible rating that indicates that the country has an extremely strong capacity to repay all its debts and related financial obligations; while the AA+ rating indicates that it has a very strong capacity to repay all its debts and related financial obligations. Fitch cited expected fiscal deterioration for the next three years and rising government debt burden that now stands at $30 Trillion as well as the recurring political gridlock in Washington over debt ceilings as the reasons for the downgrade (Fitch Ratings, 2023; and Kose et al., 2020 ).

The sovereign credit rating downgrade of the US is expected to have severe consequences on the US economy and its capital market. Evidence in the literature has affirmed this. Kaminsky, and Schmukler, (2002) argued that changes in sovereign credit ratings in countries result in severe impacts on private businesses, financial markets, and general economic conditions. In a related study, Almeida et al. (2017) confirmed that sovereign debt downgrades can have significant effects on financial markets and real economies as firms cut down on investment and further borrowing from the capital markets due to elevating cost of debt capital arising from the sovereign credit downgrade. Drago and Gallo ( 2016) adopted an event study methodology to investigate the impact of changes in sovereign credit rating on sovereign debt issuers in the Eurozone, where they demonstrated that rating downgrades and upgrades usually affect financial markets considerably as the result of the release of new information after rating change announcement and the role of rating in current financial regulation. Bayar and Kilic (2014) established a relationship between sovereign credit rating and Foreign Direct Investment in Turkey. They maintained that there was a two-way causality relationship between foreign direct investment and sovereign credit rating by Standards and Poors and Fitch, which implied that whenever there was an upgrade in the rating, the volume of foreign direct investment in the country increased and vice versa

Few connected studies have focused on the impact of sovereign credit impairments on the stock market and investigated how changes in sovereign credit ratings can affect equities in the affected country.  Kaminsky and Schmukler (2002) understudied the effect of changing sovereign credit ratings on stock markets in emerging economies and confirmed that changes in sovereign credit ratings affect equity markets in emerging economies as well as generate cross-country contagion that is capable of spreading to stock markets in the neighboring countries. They further argued that there was a bi-directional effect where credit rating upgrades usually occur during stock market rallies while credit rating downgrades take place during stock market downturns.  In another corroborative study conducted in emerging markets, Rodolfo M, (2005) supported this claim by positing that sovereign credit ratings assigned by CRAs like Moody's and Standard and Poors have a significant effect on the cross-section of domestically publicly traded stocks. Another research conducted on five Asian countries by Li, *et al* (2008) using the panel estimation method concluded that changes in sovereign wealth credit ratings announced by CRAs affected the stocks in the five Asian countries significantly. Lee, *et al* (2016) expanded the scope of their study to cover stocks in 40 countries for the period from January 1990 to March 2003 and investigated the impact of sovereign credit rating changes on stock liquidity globally. They found that changes in sovereign credit ratings significantly affected stock liquidity over the study period.  According to them, the announcement of credit downgrades has a stronger impact than credit upgrades, and the loss of investment grade strong negative effect on the stock liquidity.

Traditionally, sovereign debt impairments and rating downgrades are assumed to be associated with emerging economies that are characterized by high levels of political and economic instabilities, hence, most studies in this knowledge domain have been devoted to investigating the impact of this phenomenon on emerging market economies and their stock markets (Gadanecz & Jayaram, 2015; Abedifar, Molyneux, & Tarazi, 2013; Hou, Cheng, & Chong, 2014; Osman, Gasbarro, & Zumwalt, 2017). This stereotyping of emerging economies as the hotspot for sovereign credit downgrades and the consequent focusing of research efforts on them has been done at the expense of developed countries such as the United States as there is a scarcity of research that has examined the effects of sovereign credit downgrades on the broad US economy, let alone studies focusing on the impact of such downgrades on the US stock market (Cantor & Packer, 1994). It is interesting to note that the US in its entire history as a country has only had two sovereign credit rating downgrades- in 2011 and the current downgrade of August 2023. So, the country has traditionally enjoyed an excellent and highest possible sovereign credit rating, which might have made such studies unnecessary. However, the recent COVID-19 pandemic crisis has wreaked havoc on the US economy and forced structural shifts in the economic and political conditions, with attendant rising government debt triggering a credit rating downgrade by the CRAs (Brookings Institution, 2023; GAO, 2023; and The White House, 2022). This study investigated the impact of the recent sovereign credit downgrade on the US stock market due to the fallout of the Covid-19-driven elevated public spending and structural economic shifts using the event study methodology thus contributing to improving the available literature examining the effects of sovereign credit downgrades on the stock market within the US economic context in the post Covid-19 pandemic era.

**2.0 Research Methods and Data**

**2.1.1  Data**

This study utilized secondary financial time series as the major type of data. The variables of interest to the researcher were the stock prices of the sample of the three most capitalized companies listed on NASDAQ, which was the second largest US stock exchange. The selected companies were Apple (AAPL), Microsoft (MSFT), and Amazon (AMZN). The researcher chose these companies because they had the largest market capitalization at the time of the announcement of the US credit downgrade on the 1$^{st}$ of August 2023, hence, they were a good measure of the representation of US equities. Another variable of interest in this study was the S&P 500 daily equity market index, which was used as a broad representation of the entire US equity market. This was incorporated as a proxy to investors' sentiment to gauge how investors and the entire stock market reacted to the US credit downgrade event.

The daily stock prices of the three sampled companies and the S&P 500 equity index were collected from Yahoo Finance over two periods. The first period was 250 days commencing from the 2$^{nd}$ of November 2022 to the 10$^{th}$ of July 2023.  The second period commenced on the 2$^{nd}$  of July 2023 and ended on the 10$^{th}$ of August 2023.  The daily prices of these stocks –Apple, Microsoft, Amazon, and the S&P500 Index were transformed into daily logarithmic returns and fed into the research models as variables.

## 2.1.2 Variables

The target variables in this study were the daily logarithmic returns on Apple (AAPL), Microsoft (MSFT), Amazon (AMZN), and the logarithmic returns on the S&P 500 Index. The log returns formula below was utilized to compute these variables.

$$R_t = In(\frac{P_t}{P_{t-1}})$$

Where:

$R_t$ = Daily returns on the selected stock

$In$ = Natural logarithm of the ratio of current and previous day prices of the stock

$P_t$ = Current daily price of the stock

$P_{t-1}$ = Previous day price of the stock

The logarithmic daily returns ($R_t$) of the sampled stocks- AAPLE, MSFT, and AMZN were treated as dependent variables while the logarithmic daily returns on S&P 500 were used as independent variables. Logarithmic returns were used over simple returns because they are time additive and more normally distributed than simple returns, thus making them more amenable to statistical techniques and models that require normality of data distribution (Sharpe, 1992).

Daily log return on the sample stocks – AAPLE, MSFT, and AMZN were used as dependent variables while the daily log return on the market- S&P 500 was the independent variable. These variables were incorporated into the market models that were used to calculate the normal returns within the estimation window.

## 2.2.0 Methods

This research paper has adopted a multi-company event study as its primary methodology. The use of an event study was considered most appropriate for this research because it was the best methodology that allowed the researcher to assess the impact of changes in variables such as sovereign credit ratings on the financial market immediately after such changes were announced (Rusike and Alagidede, 2021). The significant event in this study was the Sovereign Credit Rating downgrade of the US government debt. Fitch Ratings downgraded the US debt instruments from AAA to AA+ due to the deteriorating fiscal situation, rising government debt level, and falling governance standard. This downgrade was announced on the 1st of August 2023.

Following the event study's methodology, this research was divided into two major periods called the estimation window and the event window.

### 2.2.1 Estimation Window

The estimation window was the period before the actual downgrade event occurred. The estimation window of 250 days was adopted in this study, spanning from 2nd November 2022 to 10th July 2023. The estimation window used covered a considerable amount of time to model normal returns and ensured that the estimates of model parameters were robust considering the short event window of 11 days that was adopted in this study (Xu et al., 2021 ).

### 2.2.2 Event Window

The event window used in this study comprised three sub-periods: the period shortly before the actual event (sovereign credit downgrade on 1st August, 2023), the actual event day (1st August 2023), and the period immediately after the event. The study utilized the event window of 11 days denoted as [ -5, +5], which meant 5 Days before the actual event Day (Day 0) and 5 Days after the event day. The event day was the 1st of August when Fitch Ratings announced the credit rating downgrade on the US sovereign debts. The adoption of a short event window was required since the study utilized high-frequency data (daily stock prices) that allowed rapid information diffusion, thus enabling better isolation of the immediate impact of the event on stock prices (Heston et al., 2020)

The gap called a buffer period of 10 days was allowed between the end of the estimation window and the start of the event window. This buffer period was necessary to prevent the contamination of the estimation window that might arise from any early leaks of the event information as well as mitigate market microstructure issues that are usually prevalent close to the event day.

### 2.2.3 Model Estimation and Expected Returns

The market model was used to estimate the normal returns on the stocks of the selected companies- AAPL, MSFT, and AMZN over the estimation window. Three models were specified- one model for each company stock as below.

$$R_{AAPL} = \alpha_i + \beta_i R_{mt} + \varepsilon_{it} \quad \text{...................(i)}$$

$$R_{MSFT} = \alpha_i + \beta_i R_{mt} + \varepsilon_{it} \quad \text{...................(ii)}$$

$$R_{AMZN} = \alpha_i + \beta_i R_{mt} + \varepsilon_{it} \quad \text{..................(iii)}$$

*Where:*

$R_{it}$ = estimated normal returns on AAPL, MSFT and AMZN

$R_{mt}$ = Returns on the market − S&P500 index at time t

$\alpha_i$ and $\beta_i$ = parameters to be estimated

$\varepsilon_{it}$ = error term

The above regression models were used to estimate normal returns for each stock over the 250-day estimation window before the event occurred. These models were used to estimate the normal returns for the sampled stocks- AAPL, MSFT, and AMZN in the event window.

### 2.2.4 Abnormal Returns and Cumulative Abnormal Returns

Abnormal returns were computed during the event window (22nd July 2023 to 10th August 2023) as the difference between the actual returns of each stock (AAPL, MSFT, and AMZN) for each day and the expected returns for the day. This was done using the under-listed formula.

$$AR_{it} = R_{it} - E(R_{it})$$

Where:

$A_{it}$ = Abnormal returns

$R_{it}$ = Actual daily returns on stock

$E(R_{it})$ = Expected daily returns on stock

Cumulative Abnormal returns (CAR) were computed by aggregating the above abnormal returns over the event window starting from the event day [ 0] back to 10 days before the event day and 10 days after the day – this time range was denoted as [ -10, +10 ] while the actual event day, i.e. the day of announcement of the US credit downgrade announcement was denoted as Day 0 (1st August 2023). The use of CAR in this study was informed by CAR's ability to capture the long-term effects of an event on the stock market that may not be reflected fully on the event day but manifest days or weeks after the event day. CAR enabled the researcher to examine the sustained impact of the event on the stock market beyond the event day (Keim & Vissing-Jørgensen, 2006).

The Cumulative Abnormal Returns (CAR) was determined as below.

$$CAR_{i,t} = \sum AR_{i,t}$$

Where:

$CAR_{i,t}$ = cumulative abnormal returns of each stock over the event window

$\sum AR_{i,t}$ = Summation of arbnormal returns of stocks over the event window

## 2.3 The Study Hypotheses

This study was designed to empirically test the following research hypotheses.

$H_0$: The US Sovereign Credit Downgrade does not have a significant effect on its Equity Market

$H_1$: The US Sovereign Credit Downgrade has a significant effect on its Equity Market

## 3.0 Result Presentation

The result of the Ordinary Least Square (OLS) regression models that were estimated during the event estimation window culminated in three estimated market models- one for each sample stock as below.

The equation (i) was for Apple stock, equation (ii) was for Microsoft Stock, and equation (iii) was for Amazon stock.

$$R_{AAPL} = 0.0004 + 1.3233\ R_{mt} + \varepsilon_{it} \ldots\ldots (i)$$

$$R_{MSFT} = 0.0013 + 1.3402\ R_{mt} + \varepsilon_{it} \ldots\ldots (ii)$$

$$R_{AMZN} = 0.0005 + 1.6323\ R_{mt} + \varepsilon_{it} \ldots\ldots (iii)$$

**Table 3.1: OLS Regression Results**

| Stock | Constant | Beta | p_Values | Standard Errors | R-Squared |
|---|---|---|---|---|---|
| Apple | 0.0004 | 1.3233 | 0 | 0.67 | 0.7 |
| Microsoft | 0.0013 | 1.3402 | 0 | 0.092 | 0.56 |
| Amazon | 0.0005 | 1.6323 | 0 | 0.121 | 0.52 |

Table 3.1 above presents the parameter estimates from the three market models for the sampled stocks and their key statistics. All the estimated three market models had positive constants, i.e. average daily returns for AAPL, MSFT, and AMZN were 0.0004, 0.0013, and 0.0005 respectively assuming the daily returns on the market ($R_{mt}$) which was the independent variable in these models were zero.

The estimated betas ($\beta_i$) of all the stocks- AAPL, MSFT, and AMZN were also positive and greater than 1, which indicated that the estimated daily returns on all three stocks were more volatile than the daily returns on the S&P 500 i.e., market returns during the event estimation window ( 2nd Nov. 2022 – 10th July 2023).  The daily returns on the AMZN stock were the most volatile with the highest beta of 1.6323 while the daily returns on AAPL were the least volatile with a beta of 1.3233. The estimated market models for the sampled stocks had p-Values of 0.000 for AAPL, 0.000 for MSFT, and 0.000 for AMZN (Table 3.1) and indicated that all models were statistically significant and that the daily returns on the market (S&P 500) were a great predictor of the daily returns on the individual stocks of the three sample companies.

The strong relationship between the individual stock daily returns (dependent variables) and the daily market returns -S&P 500 (independent variables) in this set of market models was evident in the high R-squared value associated with each model as shown in Table 3.1.  The estimated market model relating to AAPL stock had the highest R-squared of 0.70 which indicated that daily returns on the Market –S&P500 accounted for about 70% changes in the daily returns on AAPL stock.  This was followed by the market model of MSFT with the R-squared of 0.56 and indicated that daily market returns (S&P 500) accounted for 56% of the changes in the daily returns on MSFT stock. The market model for AMZN stock had the least R-squared of 0.52 and implied that 52% of the changes in the daily returns on this stock were explained and driven by the daily returns on the market –S&P 500.

Table 3.2 presents the daily abnormal returns ($AR_{it}$) and the cumulative abnormal returns ($CAR_{i,t}$) during the event window, i.e. from 25th July 2023 to 8th August 2023.

Table 3.2: Abnormal Returns and CAR for AAPL, MSFT, & AMZN Stocks During Event Window

| Date | Abnormal Returns_AAPL | Abnormal Returns_MSFT | Abnormal Returns_AMZN |
|---|---|---|---|
| 7/25/2023 | 0.00040 | 0.0118 | -0.0025 |
| 7/26/2023 | 0.00430 | -0.0395 | -0.0079 |
| 7/27/2023 | 0.00150 | -0.0138 | 0.0108 |
| 7/28/2023 | 0.00000 | 0.0084 | 0.0139 |
| 7/31/2023 | 0.00080 | -0.0105 | 0.0082 |
| 8/1/2023 | -0.00120 | 0.0035 | -0.0111 |
| 8/2/2023 | 0.00240 | -0.0093 | -0.0045 |
| 8/3/2023 | -0.00440 | -0.0004 | 0.0091 |
| 8/4/2023 | -0.04260 | 0.0092 | 0.0876 |
| 8/7/2023 | -0.02970 | -0.0063 | 0.0036 |
| 8/8/2023 | 0.01050 | -0.0080 | -0.0098 |
| **CAR** | **-0.058** | **-0.055** | **0.097** |

The above daily abnormal returns have been visualized in Fig.3.1 below to demonstrate the visual effect that the announcement of US sovereign credit rating downgrade by Fitch Ratings on the 1st August 2023 had on the three most capitalized company stocks on the US Stock markets.

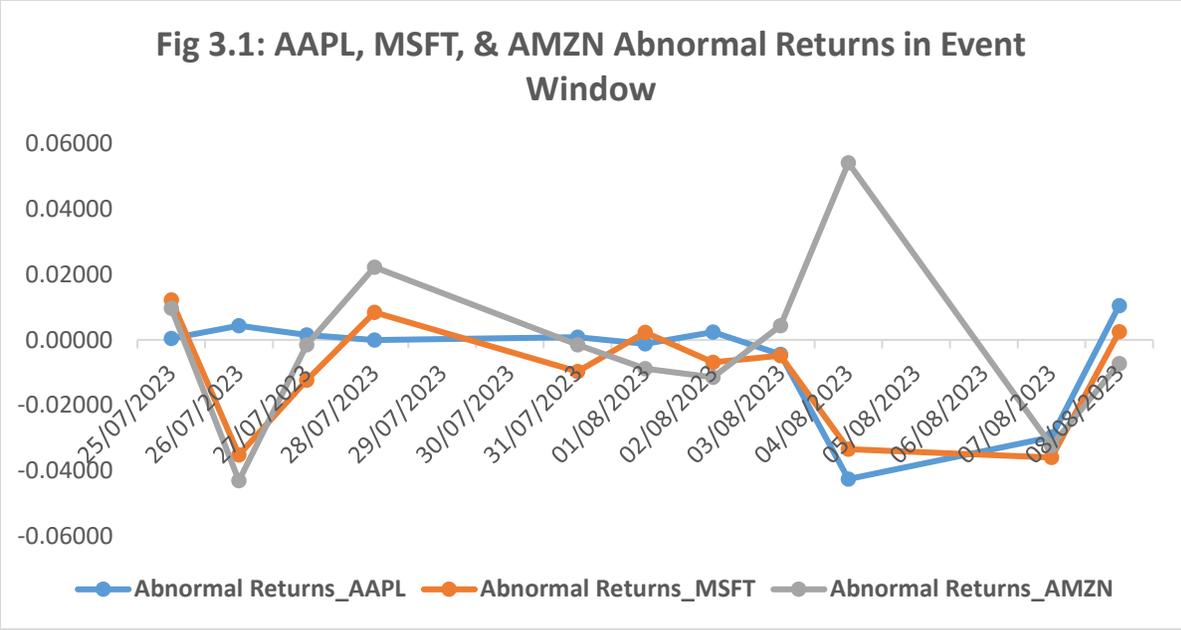

Fig 3.1: AAPL, MSFT, & AMZN Abnormal Returns in Event Window

The chart presented three lines- blue (Abnormal Returns for AAPL), orange (Abnormal Returns for MSFT), and grey (Abnormal Returns for AMZN). The chart indicated that the abnormal returns for the three stocks were both positive and negative during the event window i.e. the duration we expected the effect of the US Sovereign credit downgrade to reflect on the US stock market performance.

The day of the announcement of the credit downgrade by the credit rating agency- Fitch Rating was on the 1st of August 2023. Before this date, the daily abnormal returns on the two of the most capitalized companies- MSFT and AMZN experienced their biggest decline of over -4.0% in the pre-announcement period of the event window, precisely on 27/07/2023 while the abnormal returns of AAPL stock were stable in this period as shown in Fig.3.1 above. On the announcement day (1st of August 2023) of the credit downgrade- i.e. the event day, the abnormal returns on the three stocks behaved differently as MSFT stock's daily abnormal return gained marginally by 0.35% while AAPL and AMZN stocks' daily abnormal returns fell by -0.12% and -1.11% respectively. It is apparent that both the gains and losses on these stocks were small on the actual event day (01/08/2023) - the day of the announcement of the US sovereign credit downgrade. In the post-announcement period of the event window, the effects of the US Sovereign credit downgrade on the sampled companies' stocks became more obvious as AAPL daily abnormal returns fell the most by -4.26 % while AMZN stock posted the most significant rise; rising sharply by 8.76%. MSFT posted a moderate positive daily return of 0.97% in the post-announcement period of the event window. Overall, the daily abnormal returns of the three companies' stocks showed mixed behavior throughout the event window i.e. in some days the returns fell, and in some, they rose.

The Cumulative Abnormal Returns (CAR) of the three companies' stock presented in Table 3.2 above showed that both AAPL and MSFT had negative CARs of -0.058 and -0.055 respectively while AMZN had a positive CAR of 0.097 during the event Window. The negative CARs suggested that both AAPL and MSFT actual stock returns on average experienced unexpected declines relative to their expected returns. On the other hand, the positive CAR indicated that AMZN's actual stock returns, on average experienced an increase relative to its expected stock returns based on the market model.

Finally, to test the two research hypotheses developed earlier in the methodology section and determine the overall effect of the US sovereign credit downgrade on the equity market, a two-tailed t-test was conducted and the result was presented in Table 3.3 below.

Table 3.3: t-Test Statistics Result

| Stock | CAR | t-Value | Critical t_value | *P_value* |
|---|---|---|---|---|
| AAPL | -0.058 | -1.0952 | 2.2281 | 0.2991 |
| MSFT | -0.055 | -1.5024 | 2.2622 | 0.1672 |
| AMZN | 0.097 | 1.0980 | 2.2623 | 0.3007 |

From the above table, the t-value for AAPL stock's abnormal returns was -1.0952. This was less than the critical t-value of 2.2281 at a 95% level of significance. Also, the corresponding p-value associated with this t-value was greater than the 0.05 significance level, indicating that there was no evidence to reject the null hypothesis ($H_0$) that the US Sovereign debt credit rating downgrade had no significant effect on the US equity market. The t-value for MSFT abnormal returns was -1.5024 and less than the critical t-value of 2.2622. Its corresponding p_value was 0.1672 which was greater than the significance level of 0.05, suggesting that the null hypothesis ($H_0$) could not be rejected. Similarly, AMZN's stock abnormal returns had a t_value of 1.0980 which was less than its critical t_value of 2.2623, and a corresponding p_value of 0.3007 which was greater than the level of significance of 0.05 and indicated that there was no sufficient evidence to reject the null hypothesis ($H_0$).

## 4.0 Result Discussions

This study established that sovereign credit rating downgrade of the US sovereign debt by a credit rating agency (CRA), Fitch Ratings did not have any significant effect on the US equity markets. This finding was empirically established on the result of the two-tail t-test of the research hypotheses which showed the lack of sufficient evidence to reject the null hypothesis that the US sovereign credit rating downgrade had no significant effects on the US equity market. I could not reject this null hypothesis as there was no sufficient evidence to do so(Field, 2013). This result contradicted the expectations that the null hypothesis should have been rejected to signify that a country's sovereign credit downgrades have significant effects on its stock market. The failure to reject the null hypothesis did not mean that the US sovereign credit downgrade did not have any effect on the equity market at all. It only meant that there was no sufficient evidence at a 5% level of significance to reject the null hypothesis (Nuzzo et al., 2015). The lack of sufficient evidence could be attributed to the small sample size which restricted the statistical power of the test to detect real effects (Cohen et al., 2013). The result is very significant in that it contributed to the existing literature regarding the relationship that exists between a country's sovereign credit rating and the stock market with a focus on the US. While several studies have examined the effect of sovereign credit rating downgrade on the stock market in emerging economies which are generally perceived to have unstable economic and financial systems (Gadanecz & Jayaram, 2015; Abedifar, Molyneux, & Tarazi, 2013; Hou, Cheng, & Chong, 2014; Osman, Gasbarro, & Zumwalt, 2017), studies examining the effect of sovereign credit rating downgrade on the US stock market are grossly inadequate (Cantor & Packer, 1994) because the US has historically had only two credit rating downgrade throughout its existence as a sovereign nation, so credit rating downgrade is something very new and strange to the US financial markets.

This finding is similar to those of other studies that investigated the impact of sovereign credit downgrades on countries' equity markets and have reported mixed and insignificant results on how the downgrades affected the stock markets. Fama (1970) found that in a highly efficient market, the potential effect of the downgrade might be reflected in the stock prices even before the event day so that on the actual day of the announcement of the event, the stock market does not move significantly. Clark et al., (2018) also reported that the magnitude of the downgrade may depend on the creditworthiness of the country before the downgrade and might determine how significant the downgrade will affect the stock market. Finally, MacKinlay (1997) found that country-specific factors like existing economic conditions, political stability, and the overall health of the economy before the downgrade can influence the severity or level of impact that a credit downgrade could have on a country's stock market.

However, these findings have sharply contrasted with those reported in similar studies by Chen, et al. (2023), Barth, et al (2022), and Bekaert et al. (2021) who claimed that sovereign credit downgrades had significant adverse effects on a country's stock market.

My findings in this study have implications for both investors and financial regulators in the United States. The practical implication to the investors is the confirmation that the sovereign credit rating downgrade did not result in any significant difference between expected stock returns and actual stock returns, hence, credit downgrades do not heighten volatilities in equity market nor cause any abnormal returns to equity portfolios; at least in the short run. By leveraging this finding, investors can make more informed and confident investment decisions about their portfolio allocations during periods of economic uncertainty that are underpinned by a sovereign credit downgrade. Furthermore, financial regulators who are responsible for promoting the stability of the financial market will also be guided by the findings of this study which evidences that sovereign credit downgrades do not cause serious fluctuations in stock prices, thus they do not pose a serious threat to the broad financial market stability. This guidance is important to help regulators determine the appropriate measure of any intervention they may want to provide to support the stock market and entire financial system during periods of economic uncertainty driven by sovereign credit downgrades.

While this study provided valuable insight into the effects of sovereign credit downgrade on the US stock market, it could not be described as being exhaustive as I had some constraints that served as the limitations of the study as well as presented opportunities for future studies by other researchers. The first limitation was that the study considered a short event window of only 10 days which resulted in a small sample size for the test of the study hypotheses- a situation that could affect the statistical power of the test to detect real effects (Cohen et al., 2013). Future studies that would increase the event window to cover more days before the event and after the event, and consequently increase the sample size of the study are highly recommended. Second, the result of the test statistic could have been different if a different level of significance other than 5% significance level was adopted to test the research hypotheses. It is therefore recommended that future studies should explore the possibility of using 1% and 10% levels of significance to test the research hypotheses to determine if there would be sufficient evidence to reject the null hypothesis. Finally, this study did not consider the effect of external factors affecting the stock market during the event window, hence making it difficult to determine if other factors such as corporate earning announcements, mergers and acquisition announcements, etc had occurred during the event window and affected the behavior of the stock returns as well. Future studies that would consider external factors and isolate their impacts on the behavior of stock returns during the event window will be beneficial.

## 5.0 Conclusion

This study investigated the impact of the US Sovereign credit rating downgrade on the US stock market using the event study methodology. The result of the study was mixed and insignificant as it showed that the downgrade of the US sovereign credit rating did not have a significant effect on the US stock market.

The result of this study has significantly contributed to the existing literature on the impact of sovereign credit rating downgrades on the stock market as it has added to the insufficient amount of studies that have primarily examined the effects of sovereign credit rating downgrades on the stock markets within the US economic context. Sovereign Credit rating downgrades are not common to the US economy as bonds issued by the government have always enjoyed a long-term investment-grade credit rating of AAA (highest possible rating) until the $1^{st}$ of August 2023 when they were hit by a downgrade down to AA+ ( $2^{nd}$ highest possible rating). This result is both consistent and inconsistent with the findings of existing studies that have examined the impact of sovereign credit rating downgrades on stock markets around the World.

The limitations of this study were the small sample size, the adoption of a restricted single level of significance of 5% to test research hypotheses instead of using multiple levels of significance such as 1% and 10% to experimentally test the research hypotheses, and the failure of the study to consider external factors that might have effects on stock returns during the event window and isolate the effect of external factors on stock returns. Future studies can benefit from these by increasing the sample size during the event window by like say 100%; adopting multiple levels of significance of 1% and 10% to test research hypotheses; and considering the external factors that might occur during the event window and isolate the effects of such factors from sample stock returns. It is believed that if these limitations are remedied it could improve the result of this study and make it more reliable and generalizable.

This study has undoubtedly demonstrated that although sovereign credit downgrades might have significant adverse effects on stock markets in emerging economies as found by many studies outside the US, such results cannot be generalized and applied within the context of a developed economy with a stable and matured economic and financial system like the US as shown in the result of this study. With the evidence from this result, Investors and fund managers in the US equity markets may no longer have to panic and rush to rebalance stock investment portfolios during periods of market uncertainty characterized by sovereign credit downgrades as it does do have any significant effects on their stock portfolio returns.